\documentstyle[12pt]{article}
\begin{document}
\begin{center}
\Large \null\vskip 2cm NEW PERTURBATION THEORY IN QED

\vskip 5pt
                      G.M.Filippov
\vskip 5pt

       Cheboksary Institute Of The Moscow State Open University,
                Cheboksary, Russian Federation

\vskip 3pt

E-mail: scorp@cbx.ru

\end{center}
\vskip 8pt
\begin{quote}
 {\small \bf The perturbation theory in QED used the exact
solution with taking into account the special form of interaction
is constructed. The mean electromagnetic field of charged particle
is calculated. The possibility of elimination the problem with
ultraviolet as well as infrared divergences is shown. The
electromagnetic energy of the particle turns out to be regular and
small.}
\end{quote}
\vskip 10pt

\large \centerline {\bf 1. Introduction}
\vskip 5pt

In the conventional Dyson-Feinman approach in quantum
electrodynamics all quantities are calculated within the
perturbation  theory with using the zeroth order photon and
electron propagators (see, e.g., the famous monographs
(\cite{BD},\cite{IZ},\cite{R},\cite{W},\cite{BLP}). Fundamental
difficulty of the approach is a divergence of integrals
representing radiative corrections to the electron mass, electron
charge or power of interaction between electrons and photons. A
renormalization procedure allows us to eliminate the divergences
with the help of any form of subtracting the infinitely large
terms in the perturbation theory. Some famous physicists
(including Feinman \cite{F}, Landau and Lifshitz, Gell-Mann
\cite{BLP}) were estimating this procedure as incorrect. There is
a number of attempts to solve the problem with divergences in the
framework of conventional theory. In the last few years some works
was published where the authors find the possibility to avoid the
divergence problem as in QED as well as in the other field
theories with the help of "clothing" procedure firstly proposed in
\cite{GS} (see, e.g., \cite{HMR}, \cite{KSO}, \cite{SS},
\cite{KS}, \cite{St} and references therein). In this connection
one should note, that the "clothing" is fulfilled within the
conventional perturbation theory. On the other hand, it is
interesting to know, whether this difficulty is the property only
of the perturbation theory or of the quantum electrodynamics
itself? For answer this question one should construct any other
form of the perturbation theory, quantitative different from the
Dyson-Feinman approach. We would like to show that the appropriate
form of the theory indeed can be constructed where as infrared as
well as ultraviolet divergences turn out to be absent.  The
general scheme of the theory was described in \cite{Fil1}. The
relativistic units $(\hbar=1, c=1)$ are used throughout the paper.
The normalization volume is set equal to unity. We use the
common-famous notations of four-dimensional relativistic theory of
fields. The Greek indices run over 0,1,2,3; the Latin one - 1,2,3;
$g_{\alpha \beta}$ is the metric of the pseudo-Euclidean
space-time with the signature (1,-1,-1,-1).

\vskip 10pt

\centerline{\bf 2. Basic Definitions} \vskip 5pt

Consider the interaction of a free electron with the field of its
own radiation. Consider electron as quantum of the spinor field
with the wave operator in usual form
\begin{equation}\label{first}
{\hat\psi}=\sum\limits_{\lambda,\,\vec p}[u_{\lambda\vec
p}e^{-ipx}{\hat a}_{\lambda\vec p}+v_{\lambda\vec p}e^{ipx}{\hat
d}_{\lambda\vec p}^{\dag}],
\end{equation}
where $\vec p$ - is the momentum of the electron, $\lambda=\pm
1/2$ denotes the projection of the electron spin on the the
specific axe of quantization (down we set this one is z-axe).
Bispinors taken part in (1) are defined, e.g., in \cite{BLP},
$$
u_{\lambda\vec p}={1\over \sqrt{2\varepsilon_{\vec p}}} \;
\left(\sqrt{\varepsilon_{\vec p}+m}\;w_{\lambda}\atop
\sqrt{\varepsilon_{\vec p}-m} \; {(\vec n_{\vec p}{\vec
\sigma})}w_{\lambda}\right);
$$
\vskip 5 pt
$$
v_{\lambda\vec p}= u_{-\lambda,-\vec p}={1\over
\sqrt{2\varepsilon_{\vec p}}} \; \left(\sqrt{\varepsilon_{\vec
p}-m}\;{(\vec n_{\vec p}\vec \sigma)} w'_{\lambda}\atop
\sqrt{\varepsilon_{\vec p}+m} \; w'_{\lambda}\right).
$$
\vskip 5pt\noindent
 Here $\vec \sigma$ - is the vector of Pauli matrices,
$p^{0}=\varepsilon_{\vec p}=(p^2+m^2)^{1/2}$ - is an energy of the
electron obeying the momentum ${\vec  p}$. The unit spinors
$w_{\lambda},w'_{\lambda}$ describe the states with the spin
projections equal to $\lambda$. In the equation (1) ${\hat
a}_{\lambda\vec p}$ - is the electron destruction operator in a
state with the defined polarization, momentum and the energy
$\varepsilon_{\vec p}$,  ${\hat d}^\dagger_{\lambda\vec p}$ - is
the positron creation operator for the same state. All the
operators ${\hat a}_{\lambda\vec p}, {\hat a}^\dagger_{\lambda\vec
p}, {\hat d}_{\lambda\vec p}, {\hat d}^\dagger_{\lambda\vec p}$
obey the standard Fermi commutation relations.

  In the current density operator
$$
{\hat j}^{\mu}=e :{\hat{\bar\psi}} \gamma^{\mu}\hat\psi:,
$$
where :...: means the normal ordering, $e$ - is the electron
charge, $\hat{\bar\psi}=\hat\psi^{\dag}\gamma^{0}$ is the usual
Dirac conjugate operator. In the following we need the Fourier
representation of the current density operator which in our case
has the form
\begin{equation}\label{j}
{{\hat j}_{\vec q}}^{\mu}(t)=e \sum\limits_{\lambda,\sigma, \,\vec
p}\,[{\bar u}_{\lambda \vec p}\gamma^{\mu}u_{\sigma, \vec p+\vec
q}\; {\hat a}^\dagger_{\lambda\vec p}{\hat a}_{\sigma,\vec p+\vec
q}\; e^{i( \varepsilon_{\vec p}-\varepsilon_{\vec p+\vec q})t}+
$$
$$
{\bar u}_{\lambda,\vec p}\gamma^{\mu}v_{\sigma,-\vec p-\vec
q}{\hat a}^\dagger_{\lambda,\vec p}{\hat d}^{\dag}_{\sigma,-\vec
p-\vec q}\;e^{i(\varepsilon_{\vec p}+\varepsilon_{-\vec p-\vec
q})t}+ {\bar v}_{\lambda,\vec p}\gamma^{\mu}u_{\sigma,-\vec p+\vec
q}{\hat d}_{\lambda,\vec p}{\hat a}_{\sigma,-\vec p+\vec
q}\;e^{-i(\varepsilon_{\vec p}+\varepsilon_{-\vec p+\vec q})t}-
$$
$$
{\bar v}_{\lambda,\vec p}\gamma^{\mu}v_{\sigma,-\vec p-\vec
q}{\hat d}^{\dag}_{\sigma,\vec p-\vec q}{\hat d}_{\lambda,\vec
p}\; e^{-i(\varepsilon_{\vec p}-\varepsilon_{\vec p-\vec q})t}].
\end{equation}
Note, as a rule, the current density operator don't conserves the
spin of the particle. The conservation of the spin is possible
only at ${\vec q}=0$.

Now we need to construct the "approximate" current operators which
are to be commute if related to different times.  For the one
electron problem one of the appropriate expression is given by the
formula
\begin{equation}\label{j0}
{\hat j}_{{\vec q}(0)}^{\mu}(t)= f^{\mu}({\vec
q},t)\,{\widehat\rho}_{\vec q}.
\end{equation}
The 4-vector $f^{\mu}({\vec q},t)$ is defined below and
$$
{\hat\rho}_{\vec q} = \sum\limits_{\lambda, \vec k} {\hat
a}^\dagger_{\lambda,\vec k}{\hat a}_{\lambda,\vec k+\vec q}
$$
is the Fourier component of the time-independent electron density
operator. Note, the operators (\ref{j0}) include only diagonal
terms in polarization indices and if even refer to different
times, obey the commutation relations
\begin{equation}\label{comm}
\lbrack{\hat j}^{\mu}_{{\vec q}(0) }(t),{\hat j}^{\nu} _{{\vec
q'}(0)}(t^\prime)\rbrack_{-} = 0.
\end{equation}

Down we consider (\ref{j0}) as the zeroth order approximation to
(\ref{j}). To define the vector $f^{\mu}({\vec q},t)$ we impose
the condition: the current mean value of (\ref{j0}) is to coincide
with the mean value of (\ref{j}), namely,
\begin{equation}\label{be}
(t\vert {\hat j}^\mu_{\vec q}(t) \vert t)=f^{\mu}({\vec q},t)
(t\vert{\hat\rho}_{\vec q}\vert t).
\end{equation}
This condition ensures the been chosen $f^{\mu}({\vec q},t)$ gives
the sufficiently rapid convergence of series in the modified
perturbation theory. Obviously, the equation (\ref{be}) may be
solved only approximately because the exact vector of state is
unknown. \vskip 10 pt

\centerline {\bf 3. Solution to the Ground State Problem} \vskip
5pt

Denote the deviation of the current from its "zeroth" value as
$$
\Delta{\hat j}^{\mu}_{\vec q}(t)={\hat j}^{\mu}_{\vec q}(t)-{\hat
j}^{\mu}_{{\vec q}(0)}(t)
$$
and represent the electromagnetic interaction as a sum of two
parts,
 ${\hat H}_{int}(t)={\hat H}_{int}^
{(0)}(t)+{\hat H}_{int}^{(1)}(t)$,
where
\begin{equation}\label{int0}
{\hat H}^{(0)}_{int}(t)=e\,\int{\hat j}^{\mu}_{(0)}({\vec
r},t){\hat A}_{\mu}({\vec r},t)dV;
\end{equation}
\begin{equation}\label{int1}
{\hat H}^{(1)}_{int}(t)=e\,\int\Delta{\hat j}^{\mu}({\vec
r},t){\hat A}_{\mu}({\vec r},t)dV.
\end{equation}
The 4-vector-potential operator of electromagnetic field is
defined by the ordinary way as
\begin{equation}\label{Ab}
{\hat A}^{\mu}({\vec r},t)=\sum\limits_{\alpha,\vec q} g_q\lbrace
\hat b_{\alpha\vec q} e^{\mu}_{\alpha\vec q}e^{i{\vec q \vec r}}
+\hat b^\dagger_{\alpha\vec q} e^{\mu *}_{\alpha\vec q} e^{-i{\vec
q \vec r}}\rbrace.
\end{equation}
Here $\hat b^\dagger_{\alpha\vec q}$ and $\hat b _{\alpha\vec q}$
- are the operators of creation and destruction of photons in
states possessed the polarization $\alpha (\alpha=0,1,2,3)$,
momentum $\vec q$ and energy $\omega=q$. The coupling function
$g_q=\sqrt{2\pi/\omega}$  is defined, e.g., in \cite{BLP}. On
definition the different unit vectors of polarization
$e^{\mu}_{\alpha\vec q}$ are orthogonal to each other and obey the
normalization conditions (see, e.g.,{\cite{IZ}})
$$
g_{\mu \nu}e^{* \mu}_{\alpha\vec q}e^{\nu}_{\alpha'\vec
q}=g_{\alpha \alpha'}; \quad \sum_{\alpha}e^{* \mu}_{\alpha\vec
q}e^{\nu}_{\alpha\vec q}g_{\alpha \alpha}=g^{\mu \nu}.
$$

 The conditions of the relativistic invariance (see, e.g.,
(\cite{BD},\cite{IZ},\cite{R})) are satisfied when the photon
operators obey the commutation relations
$$
[\hat b _{\alpha\vec q},\hat b^{\dag}_{\beta \vec
q\:'}]_{-}=-g_{\alpha\beta}\cdot \Delta(\vec q - \vec q\:');
$$
$$
[\hat b _{\alpha\vec q},\hat b_{\beta \vec q\:'}]_{-}=[\hat
b^{\dag} _{\alpha\vec q},\hat b^{\dag}_{\beta \vec q\:'}]_{-}=0.
$$
Here $\Delta(\vec q - \vec q\:')$ equal to unity only if  ${\vec
q} = {\vec q\:'}$ and equal to zero in opposite case.  It is well
known that the non-usual commutation relations for the scalar
photons lead to the indefinite metric in the photon state space.
There are also some peculiarities in the definition of physical
quantities calculated below.

By virtue of (\ref{comm}), the equation
\begin{equation}\label{eq}
i{d\over d\,t} |\, t)={\hat H}_{int}^{(0)}(t) |\, t)
\end{equation}
have the exact solution expressed as the direct product of
extended coherent states \cite{Fil1}, namely,
\begin{equation}\label{vec}
|t)_0 =\prod\limits_{\alpha,\vec q}\exp\lbrace-i{\hat\chi}_
{\alpha\vec q}(t)-{\hat b}_{\alpha\vec q}{\hat Q}
^\dagger_{\alpha\vec q}(t)+{\hat b}^\dagger_{\alpha\vec q} {\hat
Q}_{\alpha\vec q}(t)\rbrace | 0).
\end{equation}
Here
\begin{equation}\label{Q}
{\hat Q}_{\alpha \vec q}(t)={\hat \rho}_{\vec q} Q_{\alpha\vec
q}(t),\quad {\hat\chi}_{\alpha\vec q}(t)={\hat \rho}_{\vec
q}^{\dag}{\hat \rho}_{\vec q}\:\chi_{\alpha\vec q}(t)
\end{equation}
and
\begin{equation}\label{Q1}
Q_{\alpha\vec q}(t)=-ig_q \int\limits_0^t dt' e^{\mu
*}_{\alpha\vec q}f_{\mu}({\vec q},t') e^{i\omega t'},
\end{equation}
\begin{equation}\label{ch}
\chi_{\alpha\vec q}(t)=-{i\over 2}\int\limits_ 0^t\lbrace{{\dot
Q}}^{*}_{\alpha\vec q}(t') {Q}_{\alpha\vec q}(t')-{Q}_{\alpha\vec
q} ^{*}(t'){{\dot Q}}_{\alpha\vec q}(t') \rbrace dt'.
\end{equation}
Within our approach an initial vector of state $|0)$ is a direct
product of the electromagnetic field vacuum state  $ | vac)$ and
the vector $ |\varphi_0)$ describing the initial state of a
particle, $\vert 0)=\vert \varphi_0)\bigotimes | vac)$; below, for
simplifications, we denote this state also as $\vert \varphi_0,
vac)$ . Assume the vector $|\varphi_0)$ is a free wave packet
which extends its width in course of time. We must take into
account that at $t>0$ in our interacted system each part of it
isn't independent and can not be described exactly. E. g., there
isn't the wave function for the electron but the mostly simplex
way to get an explicit information  about it consists in
calculation its density matrix.

\vskip 10 pt

\centerline{\bf 4. The New Perturbation Theory} \vskip 5 pt

If we neglect the corrections producing by the interaction ${\hat
H}_{int}^{(1)}$, then the equation (\ref{vec}) completely solves
the problems with calculating the physical quantities of interest.
The quantity $|Q_{\alpha\vec q}(t)|^2$ has the direct physical
meaning of the mean number of photons possessing the polarization
$\alpha$ and the momentum $\vec q$ created by the particle to the
given instant. Therefore,
\begin{equation}\label{Dk}
\Delta{\vec k}(t)=\sum_{\alpha,\vec q}{\vec q}|Q_{\alpha\vec
q}(t)|^2
\end{equation}
is the mean momentum loss and
\begin{equation}\label{DE}
\Delta E(t)=\sum_{\alpha,\vec q}\omega|Q_{\alpha\vec q}(t)|^2
\end{equation}
is the mean energy loss of the particle. Sometimes (e.g., in case
of the resting particle) the energy loss as well as the momentum
loss are zeros.

In the approximation considered we can calculate also many other
physical quantities (e.g., the Green function) for the particle
with taking into account the back influence of electromagnetic
quanta on particle's state. The corrections should be evaluated
with the help of a modified perturbation theory. To construct the
new perturbation theory with respect to the interaction  ${\hat
H}_{int}^{(1)}$ we introduce the "zeroth" order evolution operator
\begin{equation}\label{U0}
 {\hat U}_0 (t)=\exp\lbrace\sum\limits_{\alpha,\vec q} {\hat
Q}_{\alpha\vec q}(t) {\hat b}_{\alpha\vec q}^\dagger -{\hat
Q}_{\alpha\vec q}^\dagger (t) {\hat b}_{\alpha\vec q} -i
{\hat{\chi}}_{\alpha\vec q}(t)\rbrace.
\end{equation}
The vector (\ref {vec}) now can be rewritten as $\; |t)_0 = {\hat
U}_0 (t) | 0)$. Introduce the new representation of operators as
follows
\begin{equation}\label{modA}
{\tilde A}(t)={\hat U}_0^\dagger (t){\hat A} (t) {\hat U}_0 (t).
\end{equation}
The vector of state  $ | t\rangle={\hat U}_0^\dagger |t) $ in this
representation obeys the equation
\begin{equation}\label{Ne}
i {d\over dt} | t\rangle={\tilde H}_{int}^{(1)} (t) | t\rangle.
\end{equation}
The solution to this equation may be found in a rigorous way via
T-exponent
\begin{equation}\label{Te}
|t\rangle={\rm T}\exp{\left\{-i\int\limits_0^t {\tilde H}_{int}^
{(1)}(t')dt'\right\}}|0\rangle.
\end{equation}
\vskip 10pt

\centerline{\bf 5. Ultraviolet behavior of spectral functions}
\vskip 5pt

The approximate solution to the equation (\ref{be}) can be found
with the help of a self-consistent procedure, when we use as $|t)$
an approximate expression which functionally depends on
$f^{\mu}({\vec q},t)$. For the first approximation we suppose
$|t)\approx |t)_0$. Generally speaking, the quantity
$f^{\mu}({\vec q},t)$ must be chosen in a strong connection to the
particular problem. We consider now only the evolution of an
one-particle initial state on the background of photon vacuum. We
begin with evaluation of the left hand of the equation (\ref{be}).

Assume $\varphi_0$ is a Gauss wave packet
\begin{equation}\label{fi}
|\varphi_0)=(8\pi\delta^2)^{3/4}\sum_{\vec k}e^{-\delta^2 ({\vec
k}-{\vec k}_0)^2}{\hat a}^{\dag}_{\lambda',\vec k}|vac),
\end{equation}
where $\lambda'$ is a some initial polarization of the electron.
 Let us represent the equation (\ref{be}) in a more convenient
form. Using the well-known Baker-Hausdorff formula
$$
\exp({\hat A}+ {\hat B})=\exp({\hat A})\exp({\hat B})\exp(-{1/2}
[{\hat A},{\hat B}]),
$$
applicable for the case $[{\hat A},{\hat B}],{\hat A}]= [{\hat
A},{\hat B}],{\hat B}]=0$, we get
\begin{equation}\label{t0}
|t)_0=\exp\left[\sum_{\alpha,\vec q}\left(-i\chi_{\alpha\vec
q}(t)+ Q_{\alpha\vec q}(t){\hat\rho}_{\vec q}{\hat b}_{\alpha\vec
q}^\dagger - {1\over 2}|Q_{\alpha\vec q}(t)|^2\right)\right]|0).
\end{equation}
Note, operator ${\hat U}_0$ is diagonal in spin indices and
therefore in our approximation only diagonal part of ${\hat
j}^{\mu}_{\vec q}(t)$ contribute to the right hand of (\ref{be}).
The straightforward calculation gives:
\begin{equation}\label{j1}
(t|{\hat j}^{\mu}_{\vec q}(t)|t)\approx (0|{\hat U}_0^\dagger(t)
{\hat j}^{\mu}_{\vec q}(t){\hat U}_0(t)|0)\approx e\cdot
\exp\left[-\sum_{\alpha,\vec q'}|Q_{\alpha\vec
q'}(t)|^2\right]\times
$$
$$
\sum_{\lambda,\vec k} {\bar u}_{\lambda,{\vec k}-{\vec q}/2}
\gamma^{\mu}u_{\lambda,{\vec k}+{\vec q}/2}
e^{i(\varepsilon_{{\vec k}-{\vec q}/2}- \varepsilon_{{\vec
k}+{\vec q}/2})t}(0|\exp\left\{ -\sum_{\beta_1,{\vec
q}_1}Q_{\beta_1{\vec q}_1}^*(t) {\hat\rho}_{{\vec
q}_1}^\dagger{\hat b}_{\beta_1{\vec q}_1} \right\}\times
$$
$$
  {\hat a}_{\lambda,{\vec k}-{\vec q}/2}^\dagger{\hat
a}_{\lambda,{\vec k}+{\vec q}/2} \exp\left\{-\sum_{\beta_2,{\vec
q}_2} Q_{\beta_2{\vec q}_2}(t) {\hat\rho}_{{\vec q}_2}{\hat
b}_{\beta_2{\vec q}_2}^\dagger \right\}|0),
\end{equation}
or, equivalently,
\begin{equation}\label{j2}
(t|{\hat j}^{\mu}_{\vec q}(t)|t)\approx (0|{\hat U}_0^\dagger(t)
{\hat j}^{\mu}_{\vec q}(t){\hat U}_0(t)|0)=
e\cdot\exp\left[-\sum_{\alpha,\vec q'}|Q_{\alpha\vec
q'}(t)|^2\right]\times
$$
$$
\int{d^3 s\over (2\pi)^3}\sum_{\lambda, \vec p} {\bar
u}_{\lambda,{\vec p}-{\vec q}/2} \gamma^{\mu}u_{\lambda,{\vec
p}+{\vec q}/2} \exp\left[i(\varepsilon_{{\vec p}-{\vec q}/2}-
\varepsilon_{{\vec p}+{\vec q}/2})t\right]\times
$$
$$
\sum_{\vec k}e^{i({\vec k}-{\vec p})\vec s}
(0|\exp\left\{-\sum_{\beta_1,{\vec q}_1}Q_{\beta_1{\vec q}_1}^*(t)
{\hat\rho}_{{\vec q}_1}^\dagger{\hat b}_{\beta_1{\vec q}_1}
\right\}\times
$$
$$
{\hat a}_{\lambda,{\vec k}-{\vec q}/2}^\dagger{\hat
a}_{\lambda,{\vec k}+{\vec q}/2} \exp\left\{-\sum_{\beta_2,{\vec
q}_2} Q_{\beta_2{\vec q}_2}(t) {\hat\rho}_{{\vec q}_2}{\hat
b}_{\beta_2{\vec q}_2}^\dagger \right\}|0),
\end{equation}
Now, using the relation (see \cite{Fil1})
\begin{equation}\label{Fil1}
\sum_{\lambda,\vec k}e^{i{\vec k\vec x}}{\hat a}_{\lambda\vec k}
\exp\left\{-\sum_{\beta,{\vec q}} Q_{\beta {\vec q}}(t)
{\hat\rho}_{\vec q}{\hat b}_{\beta {\vec q}}^\dagger
\right\}|{\vec k}_0,vac)=e^{i{\vec k}_0 {\vec x}}|vac_p)\otimes
$$
$$
\exp\left\{-\sum_{\beta,{\vec q}} Q_{\beta{\vec q}}(t) e^{-i{\vec
q}{\vec x}}\;{\hat b}_{\beta {\vec q}}^\dagger \right\}|vac),
\end{equation}
where $|vac_p)$ is a particle vacuum state, one can obtain
\begin{equation}\label{f}
f^{\mu}({\vec q},t)=e\cdot\sum_{\vec p}{\bar u}_{\lambda',{\vec
p}-{\vec q}/2} \gamma^{\mu}u_{\lambda',{\vec p}+{\vec q}/2}
\exp\left[i(\varepsilon_{{\vec p}-{\vec q}/2}- \varepsilon_{{\vec
p}+{\vec q}/2})t\right]\times
$$
$$
\int {d^3s\over (2\pi)^3} \exp\left\{i{\vec s}({\vec k}_0-{\vec
p})- {s^2\over 8\delta^2}-\sum_{\alpha',{\vec q}\:'}
|Q_{\alpha'{\vec q}\:'}|^2 \left(1-e^{-i{\vec q}\:'{\vec
s}}\right)\right\}.
\end{equation}
Because $Q_{\alpha\vec q}$ depends on $f^{\mu}$, (\ref{f}) is
actually the integral equation for $f^{\mu}({\vec q},t)$.

Let us assume the initial width of the particle's wave packet is
large compared to the Compton wave length, i.e., $\delta\gg 1/mc$.
In this condition one can see in the right hand of (\ref{f}) the
main contribution to the sum occurs from the small neighborhood of
the point ${\vec p}={\vec k}_0$. Assuming in this neighborhood the
quantity $|{\vec p}-{\vec k}_0|$ is small and comparable with
$|\Delta\vec k|$, we can fulfill the series development in the
power of exponent in the subintegral expression in the right hand
of (\ref{f}) with taking into account only terms of the first and
the second order in $({\vec p}-{\vec k}_0)$. In this case the
equation (\ref{f}) is reduced to
\begin{equation}\label{fr}
f^{\mu}({\vec k},t)=e(2\pi\delta^2)^{3/2}\sum_{\vec p} {\bar
u}_{\lambda',{\vec p}-{\vec q}/2} \gamma^{\mu}u_{\lambda',{\vec
p}+{\vec q}/2}\times
$$
$$
 \exp\left[i(\varepsilon_{{\vec p}-{\vec q}/2}- \varepsilon_{{\vec
p}+{\vec q}/2})t-2\delta^2\left({\vec p}- {\vec k}_0+\Delta{\vec
k}(t)\right)^2\right],
\end{equation}
Consider the
non-relativistic case, when $k_{0}\ll m$, and
$\varepsilon_{\vec q /2 +\vec p}-\varepsilon_{\vec q /2 -\vec
p}\approx\vec q \vec p /\varepsilon_{\vec q /2}$. The stationary
point of exponent in (\ref {fr}) is found to be
\begin{equation}\label{fi}
 \vec p=\vec k_0-\Delta{\vec k}(t)+i{\vec q \;t\over \delta^2
\varepsilon_{\vec q/2}}\; ;\quad k_0\ll m .
\end{equation}
At sufficiently large width $\delta$ and at $q<2m $ the last
imaginary term in the first approximation can be neglected.
Moreover, because the width has the tendency to increase in course
of time (in the non-relativistic case this increasing
asymptotically develops proportional to time), the last term in
the right hand of (\ref {fi}) don't increase in course of time. We
see the behavior of the subintegral function in (\ref {fr}) is
defined mostly by the real part of the power in exponent.
Accepting this assumption, in the first approximation we can
calculate the sum involved in (\ref {fr}) in the stationary phase
approximation and get
\begin{equation}\label{fs}
f^{\mu}({\vec q},t)\approx e\cdot {\bar u}_{\lambda',{\vec
p}_{m}-{\vec q}/2} \gamma^{\mu}u_{\lambda',{\vec p}_{m}+{\vec
q}/2}\hskip 3pt e^{- i t \left(\varepsilon_+ -\varepsilon_-
\right)},
\end{equation}
where ${\vec p}_{m} = {\vec k}_0-\Delta{\vec k}({\vec k}_0,t)$,
$\varepsilon_{\pm} = \left(m^2+({{\vec p}_m} \pm {\vec q}/2)^2
\right)^{1/2}$, $\Delta{\vec k}({\vec k}_0,t)$ - is the mean
momentum loss of the particle. In this paper we take the equation
(\ref{fs}) as the basis for constructing the first simple variant
of the modified perturbation theory.

\vskip 10pt

\centerline{\bf 6. Mean Electromagnetic Field} \vskip 5 pt

   Consider the problem of corrections to the Coulomb field. The
mean vector-potential of the field is given as ${A^{\mu}}= \langle
t|{\tilde{A^{\mu}}}|t\rangle$. Calculate this vector in the zeroth
approximation setting $|t\rangle\approx|0)$. With the help of this
assumption we obtain
\begin{equation}\label{A}
{A^{\mu}}=2 e \: {\rm Re} \biggl\{i\sum\limits_{\alpha, \vec q}
g_{q}^2 g_{\alpha\alpha} e^{\mu}_{\alpha \vec q} e^{i{\vec q \vec
r} - i\omega_{\vec q} t} e^{\nu}_{\alpha \vec q}
\int\limits_{-\infty}^t e^{i{\omega}_{\vec q}t'} f_{\nu}({\vec
q},t')\, \rho_{\vec q}(t')
  \,dt'\biggr\},
\end{equation}
Note, in this expression we have taken into account the negative
eigenvalue sign for the destruction operator of the scalar photon
(see Appendix). Substituting (\ref{fs}) and taking into account
the summing rules for polarization vectors, we get
\begin{equation}\label{Amu}
{A^{\mu}}=2 e \, {\rm Re} \biggl\{i\sum\limits_{\alpha, \vec q}
g_{q}^2 e^{i{\vec q r} - i\omega t} \int\limits_{-\infty}^t \bar
u_{\lambda',{\vec p}_{m}-{\vec q}/2} \gamma^{\mu}
u_{\lambda',{\vec p}_{m}+{\vec q}/2}\times
$$
$$
e^{i{\omega}_{\vec q}t'}e^{i(\varepsilon_{-}- \varepsilon_{+})t'}
\, \rho_{\vec q}(t')\, dt'\biggr\},
\end{equation}
where the function  $\rho_{\vec  q} (t)$  is  the Fourier
component of particle's probability distribution in the rest
frame.

 For the uniformly moving particle we have ${\vec p'}={\vec k}_0$
and arrive at
\begin{equation}\label{Av}
{A^{\mu}}=2 e \, {\rm Re} \biggl\{i\sum\limits_{ \vec q} g_{q}^2
e^{i{\vec q \vec r} - i\omega t} {\bar u}_{\lambda',{\vec k}_{0}
-{\vec q}/2} \gamma^{\mu} u_{\lambda',{{\vec k}_{0}+\vec
q}/2}\times
$$
$$
\times \int\limits_{-\infty}^t e^{i{\omega}_{\vec q}t'} \exp
\{i(\varepsilon_{-}- \varepsilon_{+})t'\} \, \rho_{\vec q}(t')\,
dt'\biggr\},
\end{equation}
The potential $A^\mu$ is represented by the convolution of one for
the point  particle and the probability distribution for the
particular  state.  It is interesting to analyze the universal
case, corresponding to the point charge, for which $\rho_{\vec
q}=1$. In this case the last formula gives
\begin{equation}\label{Av1}
{A^{\mu}}=2 e \, {\rm Re} \biggl\{\sum\limits_{ \vec q} g_{q}^2
{\bar u}_{\lambda',{\vec k}_{0}-{\vec q}/2} \gamma^{\mu}
u_{\lambda',{\vec k}_{0}+{\vec q}/2} {\exp[i{\vec q \vec r} -
i(\varepsilon_{+}- \varepsilon_{-}) t]\over \omega_{\vec
q}-\varepsilon_{+}+ \varepsilon_{-}} \biggr\},
\end{equation}
or, in 3-representation,
\begin{equation}\label{Asc}
A^{0}=2e \;{\rm Re} \biggl \{\sum\limits_{\vec q} {g_{q}^2\over
\sqrt{4\varepsilon_+ \varepsilon_-}} w^{\dag}_{\lambda'} \biggl(
\sqrt{(\varepsilon_+ +m)(\varepsilon_- +m)}+
$$
$$
{\sqrt{(\varepsilon_+ -m)(\varepsilon_- -m)}\over |{\vec
k}_0-{\vec q}/2||{\vec k}_{0}+{\vec q}/2|} \bigl ( k_0^2-q^2 /4 +
i \vec \sigma \cdot[{\vec k}_0 \times \vec q] \bigr ) \biggr )
w_{\lambda'} {e^{ i\vec q \vec r-i(\varepsilon_+ -
\varepsilon_-)t}\over \omega_{\vec q}+\varepsilon_- -
\varepsilon_+} \biggr \};
\end{equation}
\begin{equation}\label{Avc}
    {\vec A}=2e \; {\rm Re} \biggl \{ \sum\limits_{\vec q}
{g_{q}^2\over \sqrt{4\varepsilon_+ \varepsilon_-}}
w^{\dag}_{\lambda'} \biggl( \vec \sigma (\vec p_+ \cdot
\vec\sigma) + (\vec p_- \cdot\vec\sigma)\vec\sigma \biggr)
w_{\lambda'} \times
$$
$$
{\exp [i\vec q \vec r-i(\varepsilon_+ - \varepsilon_-)t]\over
\omega_{\vec q}+\varepsilon_- - \varepsilon_+} \biggr \},
\end{equation}
where $\vec p_\pm=(\vec k_0\pm \vec q/2)\sqrt{(\varepsilon_\mp+m)/
(\varepsilon_\pm +m)}$.

Now first we evaluate the scalar potential for the particle found
in the rest. We substitute ${\vec k}_{0}=0$ and get
\begin{equation}\label{A0}
A^{0}=2 e \, {\rm Re} \biggl\{\sum\limits_{\vec q} g_{q}^2
u^{\dag}_{\lambda', -{\vec q}/2}  u_{\lambda', {\vec q}/2}
{\exp{i{\vec q \vec r}}\over \omega_{\vec q}} \biggr\}.
\end{equation}

Because of $u^\dag _{\lambda', -{\vec q}/2} u_{\lambda',{\vec
q}/2}= {m/\varepsilon_{\vec q /2}} $ after some algebra the
formula (\ref {A0}) gives
\begin{equation}\label{A00}
 A_0 = e \; {2\over \pi r} \int\limits_0^{2mr} K_0(\xi)\, d\xi,
\end{equation}
where $K_0(\xi)$ - is the MacDonald's function of zero order. It
is important to note that at  $r\to 0$  the potential (\ref{A00})
diverges, but only logarithmical as $ (Z m / \pi)\; \ln(e/m
r),\quad r\ll 1/m $. At $r\gg 1/m $  the potential coincides with
the Coulomb one. The interaction with the electromagnetic field
changes particle's form-factor on distances which don't exceed
some Compton wave lengths. The same behavior manifests the
potential which takes into account the radiative corrections in
the Feinman perturbation theory.

If we attempt to represent the potential (\ref{A00}) as the usual
retarded potential corresponded to any charge distribution, then
will see, that it is possible only if this distribution
sufficiently depends on the mass of the particle. The dependency
of the field on the mass can be connected to the effect of the
reaction force exited at photon radiation. This circumstance means
not the charge distribution but the sample action of
electromagnetic quanta wrapped around the charge creates the field
which we observe as the source of the Coulomb force.

Consider the selfenergy for the resting particle. The
electromagnetic contribution to it can be obtained in a usual
manner as the energy of an electric field
$$
 E = {1\over 8\pi}\int (-\nabla A_0)^2 \: dV = - {1\over 8\pi}
 \int {A_0\Delta A_0}\: dV.
$$
The integral can be evaluated exactly. First we introduce $\Phi =
r\, A_0$. Then, using the spherical symmetry and $\Delta A_0 = (1/
r) d^2 \Phi/ d r^2$,  bring  the energy to the form
$$
E = {8\over \pi^2} (e m)^2 \int\limits_0^\infty K_0^2 (2mr) \,dr =
e^2 m.
$$
In the  last  expression  the  value  of  the  integral  was  used
$$
\int\limits_0^\infty K_0^2 (\xi)\, d\xi = {\pi^2\over 4}.
$$
As we see, the electromagnetic contribution to the selfenergy of
elementary particle is comparatively small. In the case of
electron as well as proton the electromagnetic correction is equal
to the selfenergy multiplied by the fine structure constant. We
can confirm the famous result the main part of the selfenergy for
elementary particles has the non-electromagnetic nature.

Now we should make once more important notation. We can calculate
the general energy of all the photons wrapping around the particle
and found that the total its energy exactly consists with the
electromagnetic contribution to the selfenergy. This result allows
us to consider one as a new method to calculate the energy of
electromagnetic field, which don't use the explicit expressions
for the electric and magnetic fields.

 Consider the case of non-relativistic particle. We can set ${\vec
k}_0 \approx m {\dot {\vec r}}_0$, where ${\vec r}_0 (t)$ - is the
current mean coordinate of the particle,
$$
\varepsilon_+ - \varepsilon_- \approx {m \vec q {\dot {\vec
r}}_0(t)\over \varepsilon_{\vec q /2}};
$$
$$
\varepsilon_+  \varepsilon_- \approx m^2 \, (1+q^2/4 m^2).
$$
In this case we arrive at
\begin{equation}\label{As1}
    A^{0}=2e \;{\rm Re} \biggl \{\sum\limits_{\vec q}
{g_{q}^2\over \varepsilon_{q/2}} m  { \exp [i\vec q \vec
r-i(\varepsilon_+ - \varepsilon_-)t] \over \omega_{\vec
q}+\varepsilon_- - \varepsilon_+} \biggr \};
\end{equation}
and
\begin{equation}\label{Av1}
 {\vec A}= {\vec A}_s + 2e \; {\rm Re} \biggl \{ i
\sum\limits_{\vec q} {g_{q}^2\over \varepsilon_{q/2}} \biggl( \vec
p - {\vec q (\vec q \vec p)\over 4 \varepsilon_{q/2}(m+
\varepsilon_{q/2}) } \biggr)  \times
$$
$$
{\exp [i\vec q \vec r-i(\varepsilon_+ - \varepsilon_-)t]\over
\omega_{\vec q}+\varepsilon_- - \varepsilon_+} \biggr \},
\end{equation}
 Note, within the linear in $\vec k_0$ approximation  the
4-potential (\ref{As1}, \ref{Av1}) obeys the Lorentz condition
$$
{\partial A^\mu\over \partial x^\mu} = 0.
$$
With the help of simple transformations one can reduce the
potential (\ref{As1}) to the 3-dimensional integral
\begin{equation}\label{As2}
A^0 = {2\over \pi^2}\, e m  \int\limits_{-\infty}^t dt'
\int\limits_{0}^\infty {dk\over |\vec r -{m {\vec x}_0(t')\over
\varepsilon_{k/2}}|} \int\limits_{0}^\infty d\tau
\cos(k\tau)\times
$$
$$
 \biggl \{ K_0\left(2m |t-t'-\tau-|{\vec r}-{m{\vec x}_0(t')\over
 \varepsilon_{k/2}}||\right)-
$$
$$
K_0\left(2m |t-t'-\tau+|{\vec r}-{m{\vec x}_0(t')\over
\varepsilon_{k/2}}||\right)\biggr \}
\end{equation}
By setting $\vec p_m=0$ the potential (\ref {As2}) reduces to the
potential for resting particle (\ref {A00}).

Within the region $r\geq 1/2m$ the main contribution to the
potential (\ref {As2}) occurs from the small $k \leq 2m$. In this
region one can neglect all the nonlinear in $\gamma=k/2m$ terms in
the subintegral function and get
\begin{equation}\label{AK}
 A_\mu = {2\over \pi}\, Z m c^2 \int\limits_0^{\infty} d \tau
{{\dot r}_{(0)\mu}(t-\tau)\over |{\vec r}-{\vec r}_0(t-\tau)|}
\biggl[ K_0\left(2mc |c\tau-|{\vec r}-{\vec r}_0(t-\tau)||\right)-
$$
$$
K_0\left(2mc (c\tau+|{\vec r}-{\vec r}_0(t-\tau)|)\right)\biggr].
\end{equation}
Comparing this result with (\ref {As2}) we see that the previous
expression contains the additional retardation time $\tau>0$. The
physical explanation of this effect can be found on the basis of
action of the reaction force appearing at the virtual fast $k>2m$
photon radiation. At this radiation the particle get the great
(virtual) deflection from the center of the probability
distribution. This deflections display themselves in (\ref {As2})
in form of the retardation time.

At the distances $r\gg 1/2m$ the MacDonald's functions changes
very fast compared to the characteristic distance of the potential
inhomogeneity. If one replaces in (\ref{AK}) the Mac Donald's
functions by the corresponded delta-functions, the Lienard-Wichert
potentials are occur,
$$
A^\mu = e \int\limits_{-\infty}^t d t' {v^\mu \over |{\vec r -\vec
v}t'|} \bigl\{ \delta[t'-t+|{\vec r -\vec v}t'|] -
\delta[t'-t-|{\vec r - \vec v}t'|]\bigr\}.
$$

\vskip 10pt

\centerline{\bf 7. Electron Magnetic Momentum Field}
\vskip 5pt

The previous consideration of the mean electromagnetic field in
the sec.6 has shown, that the Coulomb field isn't the all
electromagnetic field of the electron found in the rest. The
additional contribution appears from the mean value of the vector
potential, which general expression is given by the formula
\begin{equation}\label{VA0}
\vec A(\vec r)=2 e \, {\rm Re} \biggl\{\sum\limits_{\vec q}
g_{q}^2 u^{\dag}_{\lambda',{-\vec q}/2} \vec\alpha
u_{\lambda',{\vec q}/2} {{\exp\{i{\vec q \vec r})}\over
\omega_{\vec q}} \biggr\}.
\end{equation}
After simple algebra one can reduce this expression to
\begin{equation}\label{VA1}
\vec A(\vec r)=[\bigtriangledown \Phi \times \hat{\vec\mu}_{el}],
\end{equation}
where
\begin{equation}\label{Phi}
    \Phi(r)={4 m \over \pi} \int\limits_0^\infty {\sin (x)
\:dx\over x \sqrt{x^{2}+{2mr}^{2}}}= {2\over \pi r}\int
\limits_0^{2mr} K_0(x)\,dx \;,
\end{equation}
$\hat{\vec\mu}_{el}={e\over m}\hat{\vec s}$ - is the electron
magnetic momentum. Apparently, the expression (\ref{VA1})
represents himself the vector potential of the electron magnetic
momentum magnetic field. At $r\to \infty \quad \Phi (r)\to 1/r$,
what is right for the point magnetic momentum. But at $r<m^{-1}$
the function $\Phi(r)$ has only the logarithmic behavior,
simulating the behavior of the main part of potential.

In the sec.2 it has been noted, that the interaction with the
photon field can produce the change in the electron spin
direction. Indeed, the explicit evaluation of matrix elements
$({\bar u}_{\lambda', -{\vec q}/2}\gamma^\mu u_{\lambda,{\vec
q}/2})$ shows that if in the mostly probably processes the
condition $q\gg \varepsilon_{\vec p}$ is fulfilled, then
independently of any initial polarization in course of evolution
the probabilities of all polarizations become approximately equal.
For more explicit proof of this conclusion one have to calculate
the above matrix elements at $\vec k_0=0$:
\begin{equation}\label{m0}
u^\dag _{\lambda', -{\vec q}/2} u_{\lambda,{\vec q}/2}= {m\over
\varepsilon_{\vec q /2}} \delta_{\lambda' \lambda} \; ;
$$
$$
u^{\dag}_{\lambda',-{\vec q}/2}{\vec \alpha} u_{\lambda,{\vec
q}/2} \cdot {\vec e}_{\beta \vec q} =i{q\over 2\varepsilon_{{\vec
q}/2}} \; w^{\dag}_{\lambda'}\; {\vec \sigma}\left[{\vec e}_{\beta
{\vec q}}\times {\vec n}_{\vec q} \right]  \; w_\lambda,
\end{equation}
where ${\vec n}_{\vec q}$ - is the unit vector directed along
$\vec q$, and ${\vec e}_{\beta {\vec q}}$ ($ \beta=1,2$) - are the
polarization vectors of transversal photons. Right hand of the
second equation in (\ref{m0}) occurs in consequence of interaction
of the electron magnetic momentum with magnetic field of radiating
photons. In our approximation, when only diagonal part of
Hamiltonian included in the ground state definition, we don't take
into account the spin change during the radiation. In the more
strict consideration one need to take into account the above
speculations and to get the regular form of the magnetic momentum
field.

\vskip 10pt

\centerline{\bf 8. Infrared asymptotics of the photon number}
\vskip 5pt

  Consider the asymptotic behavior of the mean number of quanta at
$\omega=q\to 0$. As it is known, in classical electrodynamics as
well as in the semiclassical approach there is the characteristic
behavior   $ n_{\alpha\vec q}\sim{1/\omega^3}$ at $\omega\to 0$.
In consequence of this the total number of radiated photons is
proved to be infinitely large. This result arises due to the rapid
increasing the photon number at $\omega\to 0$. Now we note that
taking into account the reactive radiation forces remove the
singular behavior of the total photon number. As it was shown in
\cite{Fil2}, the number of radiating photons in collision between
the moving particle and a resting scatterer is given by the
expression
\begin{equation}\label{n}
 n_{\alpha\vec q}(\infty)=e^2g_q^2 \hskip 4pt\vrule width 0.4pt
height 15pt depth 15pt\hskip 4pt {{\vec e}_{\alpha\vec q}{\vec
v}_2\over\omega-{\vec q}{\vec v}_2+\Delta}- {{\vec e}_{\alpha\vec
q}{\vec v}_1\over\omega-{\vec q} {\vec v}_1+\Delta}\hskip
4pt{\vrule width 0.4pt height 15pt depth 15pt \hskip 4pt}^2,
\end{equation}
where ${\vec v_1}$ is assuming the velocity before and ${\vec
v}_2$ - after collision, $|{\vec v}_1-{\vec v}_2|\ll v_1$,
\begin{equation}\label{Del}
\Delta=2 e^2 \sum\limits_{{\vec q}\:'} g^2_{q'}{{\vec p}_0^2 -
({\vec q}\:'{\vec p}_0)^2/{q'}^2 \over \varepsilon_- \varepsilon_+
(\omega\:'-\varepsilon_+ + \varepsilon_- + \Delta)}.
\end{equation}
Here ${\vec p}_0\approx m{\vec v}_1$. In the non-relativistic
limit $v_0\ll c$ from (\ref{Del}) we get $\Delta\approx (4/3) e^2
m v^2_1$. The formula (\ref{n}) doesn't contain the infrared
singularity. The deviation from the law  $ n_{\alpha\vec q}\sim
1/\omega^3$  at
 $\omega\to 0$ begins at $\omega\sim\Delta$. This energy is the
less the less the energy of relative motion of the electron and
the scatterer. \vskip 10pt

\centerline{\bf 8. Conclusion} \vskip 5pt

We see, that problems with divergences in the QED are more the
problems of the perturbation theory than the QED itself. The
divergences characteristic for the Feinman-Dyson perturbation
theory are eliminated due to the physical effects: i) displacement
of the particle as the reaction on the own radiation, ii) spin
overturn processes due to radiation. In the Feinman approach the
commonly used zeroth order propagators don't take into account the
electromagnetic interaction at all. Apparently, if we use the more
correct propagators within the new perturbation theory (which is
equivalent to the prior "clothing" of particles), the infrared as
well as ultraviolet divergences at the calculation of physical
quantities of interest will be eliminated.

\vskip 2 cm

\noindent
 Appendix
 \vskip 1cm \noindent

 {\bf  Coherent States for the Scalar Photon Field}

Because of the non-usual commutation relations for scalar photons,
there are some peculiarities in formulating the coherent states in
this case. Below we omit index $\vec q$ in notations of operators
${\hat a}_{0,\vec q}$ for convenience. To construct the coherent
states for the scalar photon field, applicable within the Lorentz
gauge and Gupta-Bleuer formalism, we have first to define the
Hamiltonian for the free scalar photons. In opposite to the
transversal and longitudinal photons it contains the negative sign
in front of the usual expression (see, e.g., \cite{R}).
$$
{\hat H}_{0}=-\omega {\hat b}^{\dag}_{0}{\hat b}_{0}.
$$
Consider the Schroedinger equation for the forced oscillator
\begin{equation}\label{ECS}
i{d\over dt}|t)=[-\omega {\hat b}^{\dag}_{0}{\hat
b}_{0}+\alpha(t){\hat b}^{\dag}_{0}+\alpha^{*}(t){\hat b}_{0}]|t).
\end{equation}
The vector of state in the interaction representation obeys the
equation
\begin{equation}\label{ECSI}
i{d\over dt}|t\rangle=[\alpha(t){\tilde
b}^{\dag}_{0}(t)+\alpha^{*}(t){\tilde b}_{0}(t)]|t\rangle.
\end{equation}
Here the interaction representation of operators don't differ from
usual, for example: $ {\tilde b}_{0}(t)={\hat b}_{0} e^{-i\omega
t}$.

Consider the usual definition of coherent states via D-operator
(see, e.g., \cite{P})
\begin{equation}\label{CS}
|Q_{0})={\hat D}(Q_{0})|vac)=\exp{(Q_{0}{\hat
b}^{\dag}_{0}-Q^{*}_{0}{\hat b}_{0})}|vac).
\end{equation}
Taking into account ${\hat b}_{0}|vac)=0$, we get
\begin{equation}\label{CS1}
{\hat
D}(Q_{0})|vac)=e^{|Q_{0}|^{2}/2}\cdot\sum\limits_{n=0}^{\infty}{Q_{0}^{n}\over
n!}({\hat b}_{0}^{\dag})^{n}|vac).
\end{equation}
As we see, the sign in the power of the first exponent in the
right hand of this equation is opposite compared to the ordinary
coherent state. The coherent state (\ref{CS1}) is the eigenvector
of the annihilation operator but its eigenvalue has the different
sign:
$$
{\hat b}_{0}|Q_{0})=-Q_{0}|Q_{0}).
$$
The solution to the equation (\ref{ECSI}) can be simply obtained
with the help of famous Feinman-Schwinger approach (see, e.g., in
{\cite{P}}) in the form $|t)=e^{-is_{0}}|Q_{0})$. Apply the simple
algebra, we get
\begin{equation}
\label{SECSI} |t\rangle=\exp\{{{i\over
2}\int\limits_{t_{0}}^{t}\Im[{\dot
Q}^{*}_{0}}(t')Q_{0}(t')]dt'\}\cdot{\hat D}(Q_{0}(t))|vac),
\end{equation}
where $t_{0}$ - is an instant, when we know the initial vacuum
state of the photon field,
$$
Q_{0}(t)=-i\int\limits_{t_{0}}^{t}\alpha (t')e^{i\omega t'}dt'.
$$

\noindent {\bf Acknowledgement}
\noindent\vskip 5mm
 I am thankful to Dr. V.Yu. Korda for drawing
my attention to refs. \cite{GS} -- \cite{St} and related works.

\end{document}